\documentstyle[a4wide,epsf,11pt,titlepage]{article}
\newcounter{nref}
\newcommand{\bbib}{%
  \renewcommand{\refname}{\large\bf References}%
  \begin{thebibliography}{99}%
  \setcounter{enumiv}{\thenref}}
\newcommand{\ebib}{%
  \end{thebibliography}%
  \setcounter{nref}{\arabic{enumiv}}}
\newcommand{\head}[3]{%
  \setcounter{nref}{0}%
  \thispagestyle{empty}%
  \section*{\LARGE\bf #1}%
  \stepcounter{section}%
  \addcontentsline{toc}{section}{#1}%
  \large\itshape%
  #2\\\vspace{0.1pt}\\%
  #3%
  \normalsize\upshape%
  \bigskip}
\def\be{\begin{equation}}
\def\ee{\end{equation}}
\def\ggg{G}
\begin{document}
\head{\sc Hadronic Interactions for High Energy Cosmic Ray Air Showers}
{H.J. DRESCHER$^1$, M. HLADIK$^1$, S. OSTAPCHENKO$^{2,1}$, K. WERNER$^{1*}$}
{$^1$ SUBATECH, Universit\'e de Nantes -- IN2P3/CNRS -- EMN, Nantes, France\\
$^2$ Moscow State University, Institute of Nuclear Physics, Moscow, Russia\\ \\
$^*$ invited speaker, ``Physics at Cosmic Accelerators'', Burg Liebenzell, Germany, \\
\hspace*{5cm} April 1-5, 1998
}

\subsection{Introduction}

Understanding hadronic interactions at high energies represents a very important
ingredient for modeling high energy cosmic ray air showers. In this paper, we 
present a new approach to simulate  hadronic interactions, including 
hadron-hadron, hadron-nucleus, and nucleus-nucleus scattering,
in the energy range roughly between  $10^{11}$ and $10^{17}$ eV. 
Such collisions are very complex, being composed of many 
components, and therefore some strategy is needed to construct a reliable 
model. The central point of our approach is the hypothesis, that the behavior 
of high energy interactions is universal (universality hypothesis). So, for 
example, the hadronization of partons in nuclear interactions follows the same 
rules as the one in electron-positron annihilation ; the radiation of off-shell 
partons in hadronic  collisions is based on the same principles as the one in 
deep inelastic scattering. We construct a model for hadronic interactions in a 
modular fashion. The individual modules, based on the universality hypothesis, 
are identified as building blocks for more elementary interactions (like
$e^+e^-$, lepton-proton), and can therefore be studied in a much simpler 
context. With these building blocks under control, we can provide a quite 
reliable model for nucleus-nucleus, hadron-nucleus and hadron-hadron scattering,
providing in particular very useful tests for the  complicated numerical
procedures using Monte Carlo techniques.

\subsection{The Universality Hypothesis}

Generalizing proton-proton interactions,
the structure of nucleus-nucleus scattering should be as follows:
there are elementary  inelastic interactions between individual 
nucleons, realized by partonic  ``half-ladders'', where the same nucleon may 
participate in several of these elementary interactions. 
Also elastic scatterings are possible, represented by parton ladders. 
Although such 
diagrams can be calculated in the framework of perturbative QCD, there 
are quite a few problems : important cut--offs have to be chosen, one has to 
choose the appropriate evolution variables, one may question the validity of the 
``leading logarithmic approximation'', the coupling of the parton ladder to the nucleon is 
not known, the hadronization procedure is not calculable from first principles 
and so on. So there are still many unknowns, and a more detailed study is 
needed. 

Our starting point is the universality-hypothesis, saying that 
{\it the behavior of high-energy interactions is universal}.
In this case all the details of nuclear interactions can be determined by studying 
simple systems in connection with using a modular structure for modeling 
nuclear scattering. One might think of proton-proton scattering representing a 
 simple system, but this is already quite complicated considering the fact  that 
we have in general already several elementary interactions
. 
It would be desirable to study just one elementary interaction, which we refer 
to as  ``semihard Pomeron'', which will be done in the next section.

\subsection{The semihard Pomeron}

In order to investigate the semihard Pomeron, we turn to an even simpler 
system, namely lepton-nucleon scattering. 
A photon is exchanged between the lepton and a quark of the 
proton, where this quark represents the last one in a ``cascade'' of partons 
emitted from the nucleon. The squared diagram 
represents a parton ladder.
In the leading logarithmic approximation (LLA), the virtualities of the
partons are 
ordered such that the largest one is close to the photon \cite{rey81,alt82}. 
If we compare with proton-proton scattering, we have ordering 
from both sides with the largest virtuality in the middle, so in some sense the 
hadronic part of the lepton-proton diagram represents half of the elementary 
proton-proton diagram, and should therefore be studied first. In fact such 
statements are to some extent commonly accepted, but not carried through 
rigorously in the sense that also for example the hadronization of these two 
processes is related.

But first we investigate the so-called structure function $F_2$, related to the 
lepton-proton cross section via \cite{ell96}
\be
{d\sigma\over dx\,dQ^2}=L(x,Q^2)\,F_2(x,Q^2)
\label{e1}
\ee
with a calculable factor $L$. 
The variable $Q^2$ represents the absolute value of the photon virtuality, and $x$ is the
Bjorken-$x$ variable.
$F_2$ represents the hadronic part of the diagram, and is, using eq. (\ref{e1}), 
measurable. In lowest order and considering only leading logarithms of $Q^2$, we find
\be
F_2(x,Q^2)=\sum_j e_j^2\,x\,f^j(x,Q^2)
\ee
with
\be
f^j(x,Q^2)=f^j(x,Q_0^2)+\sum_{ij}\int_x^1{d\xi\over\xi}
\int_{Q_0^2}^{Q^2}{dQ'^2\over Q'^2}\,f^i(\xi,Q'^2)\,{\alpha_s\over
2\pi}\,P_i^j({x\over \xi}).
\label{e2}
\ee
Iterating this equation obviously represents a parton ladder with ordered 
virtualities, coupled in a nonperturbative way to the nucleon 
.
To account for the perturbative part, we introduce 
a so-called QCD evolution function 
$E^{ij}_{\rm QCD}(Q_0^2,Q_1^2,x)$, representing the evolution of a parton cascade from
scale $Q_0^2$ to
$Q_1^2$.
This function is calculated in an iterative way based on eq. (\ref{e2}).
Next we have to determine the $x$-distribution of the first parton of the 
ladder. 
We consider a Pomeron contribution
\be
\varphi^i_{\rm I\!P}(x)=C_{\rm I\!P}\otimes E^i_{\rm soft\,I\!P}
\ee
and a Reggeon contribution
\be
\varphi^i_{\rm I\!R}(x)=C_{\rm I\!R}\otimes E^i_{\rm soft\,I\!R},
\ee
with$E^i_{\rm soft\,I\!P}$ and $E^i_{\rm soft\,I\!R}$
representing a soft Pomeron and a Reggeon respectively,
and $C$ is the Pomeron/Reggeon-nucleon coupling \cite{wer97}.
The sum of these two contributions is the 
total initial distribution $\varphi^i(x)$.
The distribution at scale $Q^2$ is 
\be
f^j=\sum_i\varphi^i\otimes E_{\rm QCD}^{ij}
\ee
The structure function is then calculated as 
\be
F_2(x,Q^2)=\sum_je_j^2\,x\,f^j(x,Q^2).
\ee
For $Q = Q_0$, the I$\!$P-contribution is a function which peaks at very small values 
of $x$ and then decreases monotonically towards zero for $x = 1$, the I$\!$R-contribution on the 
other hand has a maximum at large values of $x$ and goes towards zero for small 
values of $x$. The precise form of $f$ depends crucially on the 
exponent for the Pomeron-nucleon coupling, and we find a good
agreement 
for $\beta_{\rm I\!P}={1\over 2}$
.

We are now in a position to write down the expression $\ggg_{\rm semi}$ for a {\it cut
semihard Pomeron}, representing an elementary inelastic interaction in $pp$
scattering. We can divide the corresponding  diagram 
into three parts
.
We have the process involving the highest parton virtuality in the middle, and 
the upper and lower part representing each an ordered parton ladder
coupled to 
the nucleon. According to the universality hypothesis, the two latter parts are 
known from studying deep inelastic scattering, representing each the hadronic part of 
the DIS diagram. So we get,
for given impact parameter $b$
and given energy squared $s$,
\be
\ggg_{\rm semi}=\sum_{ij}\int d\xi^+d\xi^-dQ^2\,f^i(\xi^+,Q^2)\,f^j(\xi^-,Q^2)
\,{d\sigma^{ij}_{\rm Born}\over dQ^2}(\xi^+\xi^-s,Q^2).
\ee
In addition to the semihard Pomeron, one has to consider the expression
representing the soft Pomeron \cite{wer93}.
The latter one,  $\ggg_{\rm soft}$, is the Fourier 
transform of a Regge pole amplitude $A\sim s^{\alpha(t)}$. 
So an elementary 
inelastic intraction in an energy range of say $10$ - $10^4$ GeV is therefore written as
\be
\ggg_{\rm tot}=\ggg_{\rm semi}+\ggg_{\rm soft}.
\ee

Up to this point, we are able to calculate cross sections, namely $F_2$, which is
essentially the
photon-proton cross section, in case of lepton-proton scattering, and the jet cross
section, which  can be onbtained by integrating $G_{\rm semi}$ over impact parameter, in
case of proton-proton scattering.

\subsection{Hadron Production}
 
As discussed in the last chapter, there exist observables (cross sections)
which can be calculated without detailed knowledge about hadron production,
but our main goal is to calculate particle production.
Using the Monte Carlo technique,
this amounts to generating first parton configurations, and then, in a second step, hadron
configurations. 

Let us start again with the case of lepton-nucleon scattering.
We generate a parton configuration, based on the expression for $F_2$ as dicussed in  the
previous chapter.
The next step consists of generating with certain probabilities hadron 
configurations, starting from a given parton configuration. We cannot calculate 
those probabilities  within QCD, so we simply provide a recipe, the so-called 
string model. The first step consists of mapping a partonic configuration 
into a string 
configuration. For this purpose, we use the colour representation of the parton 
configuration: a quark is represented by a colour line, a gluon by a colour-anticolour 
pair
.
One then follows the colour flow starting from a quark via gluons, as 
intermediate steps, till one finds an antiquark. The corresponding sequences
\be
q - g_1 - g_2 \ldots - g_n - \bar q
\ee
are identified with kinky strings, where the gluons represent the kinks. Such a
string 
decays into hadron configurations with the corresponding probabilities given in 
the framework of the theory of classical relativistic strings.

The above discussion of how to generate parton and hadron configurations is not 
yet complete : the emitted partons are in general off--shell and can therefore 
radiate further partons. This so called timelike radiation is taken into account 
using standard techniques. The mapping of parton to hadron configurations still 
works the same way as discussed above.

In case of  proton-proton interactions one generates 
parton configurations based on the expression for $G_{\rm semi}$ as discussed in the
last chapter.
Hadron configurations are generated according to the same principles as for lepton
scattering.
Actually,
our treatment of generating parton configurations for an elementary pp 
interaction is absolutely compatible with deep inelastic scattering, it is based on 
the same building blocks, in particular on the evolution functions.

Our procedure can now be used to treat the case of many semihard Pomerons, 
for $p-p$ as well as nuclear scattering. A detailed discussion will be given in a 
future publication.

\subsection{Acknowledgements}

This work has been funded in part by the IN2P3/CNRS (PICS 580)
and the Russian Foundation of Fundamental Research (RFFI-98-02-22024). 
We thank the Institute for Nuclear Theory at the University of Washington for
its hospitality and for partial support during the completion of this work.

\end{document}